\documentclass[11pt,a4paper]{article}

\usepackage[utf8]{inputenc}
\usepackage[T1]{fontenc}
\usepackage{amsmath,amssymb}
\usepackage{bm}
\usepackage{graphicx}
\usepackage{booktabs}
\usepackage{hyperref}
\usepackage{url}
\usepackage[numbers,sort&compress]{natbib}
\usepackage{authblk}
\usepackage[margin=2.5cm]{geometry}
\usepackage{caption}
\usepackage{float}
\usepackage{enumitem}
\usepackage{xcolor}

\hypersetup{
  colorlinks=true,
  linkcolor=blue,
  citecolor=blue,
  urlcolor=blue
}

\newcommand{\Us}{U^{*}}
\newcommand{\UsF}{U^{*}_{F}}
\newcommand{\MR}{M_{R}}
\newcommand{\vect}[1]{\bm{#1}}
\newcommand{\doi}[1]{\href{https://doi.org/#1}{doi:#1}}
\newcommand{\orcid}[1]{\textsuperscript{\href{https://orcid.org/#1}{\scriptsize ORCID}}}

\title{\textbf{Rigid-Body Support Motion and Post-Flutter Piezoelectric Energy
Harvesting from a Pitch-Plunge-Flap Aerofoil}}

\author[1]{Nikolaos~D.~Tantaroudas\orcid{0000-0001-6727-1014}\thanks{Corresponding
author: \texttt{nikolaos.tantaroudas@iccs.gr}.}}
\author[2]{Ilias~Karachalios\orcid{0009-0000-8581-8335}}
\author[3]{Andrew~J.~McCracken}

\affil[1]{Institute of Communications and Computer Systems (ICCS), National
Technical University of Athens, Iroon Polytechniou 9, 15773 Zografou, Greece}
\affil[2]{Department of Environmental Sciences, University of Thessaly, Gaiopolis
Campus, Larissa-Trikala Ring Road, 41500 Larissa, Greece}
\affil[3]{DASKALOS APPS, 183 Rue de l'Abb\'{e} Griffon, 01960 P\'{e}ronnas,
France}

\date{}

\begin{document}

\maketitle

% ============================================================
\begin{abstract}
A piezoelectric transducer on a pitch-plunge aerofoil with a finite-mass trailing-edge flap
turns the limit cycle that follows flutter into electrical power. The design findings for such
a harvester have so far been obtained with the section's suspension reacting against rigid
ground, which is what a wind-tunnel mounting provides. Here
a 3-dof aerofoil with a transducer is carried by a
fuselage mass that is itself free to translate, so that the coupled system is free-free and
the wing's plunge is measured relative to a support that moves. The support degree of freedom
is destabilising on its own, softening the flutter mode by taking a share of it in antiphase
with the wing's plunge, and it acts on the three transducer mountings unequally, amplifying
the plunge mounting's destabilisation, leaving the largest pitch-mounted stabilisation
untouched while halving it where the electrical time constant meets the flutter frequency, and
leaving the flap mounting alone. The optimum coupling and load resistance do not move, so the
matching rules of the rigidly supported section carry over unchanged. The harvested power
rises with the support mass at a fixed flow speed for every mounting, and that rise is shown
to be an artefact of the fixed-speed convention rather than a property of the cycle. Measured
instead at a fixed margin above each system's own flutter boundary, which divides out the
boundary's motion, the plunge mounting keeps a real gain while the pitch mounting loses power,
so the apparent benefit of a soft support to a pitch-mounted harvester is entirely the
boundary coming down to meet the flow speed. A harvester result quoted at a fixed speed and
read as a property of the limit cycle is therefore wrong by the whole of that difference.
\end{abstract}

\noindent\textbf{Keywords:} aeroelasticity, energy harvesting, piezoelectric
transduction, limit-cycle oscillation, flutter, free-free systems, unsteady
aerodynamics

% ============================================================
\section{Introduction}
\label{sec:intro}

Flutter is normally a boundary to be designed away from, and the limit-cycle oscillations that
follow it are a fatigue problem. The same instability is also a sustained conversion of flow
energy into structural motion, and a piezoelectric transducer embedded in the structure
converts part of that motion into electrical power. Erturk and co-workers generated
$10.7$~mW at the linear flutter speed of a piezoaeroelastic airfoil and noted that the
transduction itself displaces the flutter boundary~\citep{erturk2010}. Because a linear system
at its boundary has no bounded amplitude, practical devices operate post-critically on a cycle
set by a structural or aerodynamic nonlinearity. Power has been extracted from the limit
cycles of a flexible beam in axial flow~\citep{dunnmon2011}, combining free play with cubic
hardening widens the range of speeds over which a bounded amplitude persists~\citep{sousa2011},
the piezoaeroelastic harvester has been modelled~\citep{abdelkefi2011}, piezoelectric and
electromagnetic transduction have been compared on airfoil-based
devices~\citep{demarqui2013}, and a flutter-driven harvester has been built and
tested~\citep{bryant2011}. The source considered here is a deterministic limit cycle rather
than broadband random excitation~\citep{adhikari2009}. The field is reviewed
in~\citep{abdelkefi2016}.

A companion paper~\citep{tantaroudas2026harvest}, arXiv:2609.12788, studied such a harvester on
a three-degree-of-freedom aerofoil, a pitch-plunge section with a finite-mass trailing-edge
flap and unsteady strip theory, and found that the degree of freedom carrying the transducer
sets both the sign of the shift of the flutter boundary and the harvested power. A second
paper~\citep{tantaroudas2026cfd} repeated those test cases with the aerodynamics replaced by
the compressible Euler equations and asked which of the findings survive a higher-fidelity
flow, finding that strip theory under-predicts the harvested power of the compressible section
and overstates the transducer's effect on stability where the electrical time constant meets
the flutter frequency.

Both of those papers, and the aeroelastic harvesting literature more generally, mount the
section's suspension on rigid ground. That is what a wind-tunnel rig provides, and it is the
right model of one. It is not what a flying vehicle provides. On an aircraft the wing's
bending spring reacts against a fuselage of finite mass which is free to move, the coupled
system is free-free rather than grounded, and the wing's plunge is a displacement relative to a
support that is itself displacing. The question this paper asks is whether the design findings
of the rigidly supported section survive that change, and in particular whether the ranking of
the transducer mountings, the matching of the electrical load and the harvested power at the
optimum are properties of the section or of its mounting.

The present work therefore keeps everything of the companion paper except the ground. The same
section, the same aerodynamics, the same transducer and the same test cases are carried by a
fuselage mass $M_f$ free to translate vertically, giving a four-degree-of-freedom structure
parameterised by the single ratio $\MR = m/M_f$ of the wing section's mass to the fuselage's.

The work builds on a line of nonlinear reduced-order modelling developed by the authors, whose
Taylor-expansion residual formulation with biorthonormal eigenvector projection was introduced
for gust loads analysis~\citep{daronch2013gust} and control
applications~\citep{daronch2013control}, and extended to adaptive aeroelastic
control~\citep{tantaroudas2014adaptive}, to flutter suppression validated from simulation to
wind-tunnel testing~\citep{daronch2014flutter,papatheou2013}, to very flexible
aircraft~\citep{tantaroudas2015vfa,tantaroudas2017vfa} and to rapid worst-case gust
identification~\citep{tantaroudas2026gust}. How much the level of aerodynamic modelling changes
such a reduced prediction was examined for a manoeuvring aircraft
in~\citep{daronch2014aero}. The pitch-plunge section with a
finite-mass trailing-edge flap used here is the testbed characterised experimentally and
numerically in~\citep{fichera2014}, and the free-free coupling between a wing's elastic
degrees of freedom and a body's rigid-body ones is the same coupling that makes
aeroelastic-flight-dynamic models of flexible aircraft distinctive~\citep{tantaroudas2026coupled}.

Section~\ref{sec:model} states the model and the two facts that assemble it,
Section~\ref{sec:impl} the implementation and the five checks it was put through,
Section~\ref{sec:flutter} the linear flutter boundary, Section~\ref{sec:power} the limit
cycle and the harvested power and Section~\ref{sec:rom} the reduced model.
Section~\ref{sec:discussion} states which of the companion paper's conclusions survive.

% ============================================================
\section{Electro-aeroelastic model}
\label{sec:model}

\subsection{Structure}

\begin{figure}[H]
\centering
\includegraphics[width=0.78\textwidth]{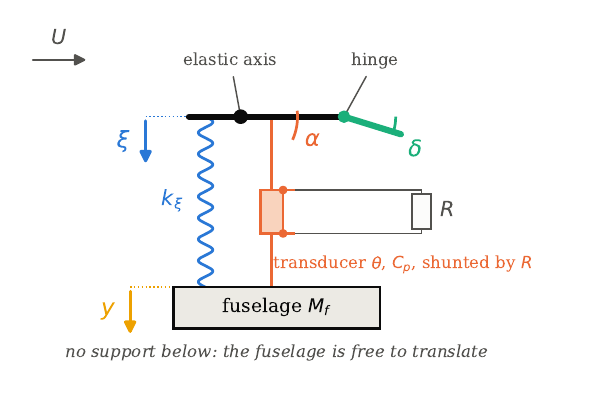}
\caption{The four-degree-of-freedom section. The plunge spring and the transducer act between
the wing and the fuselage, so $\xi$ is a relative displacement and $y$ an absolute one.}
\label{fig:schematic}
\end{figure}

The testbed is the companion paper's two-dimensional section, free to plunge, to pitch about an
elastic axis and to rotate a finite-mass trailing-edge flap about its hinge, with one addition.
Its plunge spring does not react against ground. It reacts against a fuselage of mass $M_f$,
free to translate vertically and held by nothing (Figure~\ref{fig:schematic}). Lengths are
scaled by the semi-chord $b$, time as $\tau = Ut/b$ and flow speed as the reduced velocity
$\Us = U/(b\omega_\alpha)$. The generalised coordinates are
$\vect{q} = (\xi,\alpha,\delta,y)^{\mathsf T}$, with $\xi = h/b$ the wing's plunge
\emph{relative to the fuselage}, positive downward, $\alpha$ the pitch, positive nose-up,
$\delta$ the flap deflection, positive trailing-edge-down, and $y = Y/b$ the fuselage's own
translation. Primes denote $\mathrm{d}/\mathrm{d}\tau$. The wing's absolute plunge is
$\xi + y$, and that single sentence is the model.

The mounting enters through exactly one dimensionless group, the ratio of the wing section's
mass to the fuselage's,
\begin{equation}
\MR = \frac{m}{M_f} = \left(\frac{\omega_y}{\omega}\right)^{2},
\label{eq:mr}
\end{equation}
with $\omega_y$ the frequency of the fuselage on the wing's plunge spring and $\omega$ the
plunge-to-pitch frequency ratio, both referred to $\omega_\alpha$. $\MR = 0$ is a fuselage of
infinite mass, which is the companion paper's rigidly supported section, and $\MR = 1$ is a
fuselage as heavy as the wing section it carries.

The equations of motion are
\begin{equation}
\vect{M}\vect{q}'' + \vect{C}\vect{q}' + \vect{K}\vect{q}
+ \vect{f}_{\mathrm{nl}}(\vect{q}) + \vect{W}\vect{w}
+ \kappa\,\vect{e}_{j}\,v = \vect{0},
\label{eq:struct}
\end{equation}
with $\vect{w}$ the eight aerodynamic lag states of Section~\ref{sec:aero} and the last term
the electromechanical back-coupling of Section~\ref{sec:piezo}. Two facts assemble the
$4\times4$ operators from the companion paper's $3\times3$ ones, and they pull in opposite
directions.

\textbf{The inertia is mapped on both sides.} The wing's kinetic energy depends on $y$ only
through the combination $\xi + y$, and so does the aerodynamic load, since the air sees the
absolute motion of the section and nothing else. The generalised force on $y$ therefore equals
the generalised force on $\xi$. Writing the motion map
\begin{equation}
\vect{T} =
\begin{bmatrix} 1 & 0 & 0 & 1 \\ 0 & 1 & 0 & 0 \\ 0 & 0 & 1 & 0 \end{bmatrix},
\qquad
\vect{T}\vect{q} = (\xi + y,\ \alpha,\ \delta)^{\mathsf T},
\label{eq:Tmap}
\end{equation}
every aerodynamic operator $\vect{A}_3$ of the companion paper and its structural mass become
$\vect{R}\vect{A}_3\vect{T}$, where $\vect{R}$ is $\vect{T}^{\mathsf T}$ with its fuselage row
scaled by $\MR$, and each lag state is driven by $\vect{T}\vect{q}$ rather than by $\vect{q}$.
The scaling in $\vect{R}$ is the fuselage equation's own normalisation, that equation being
divided through by $M_f b$ where the wing's are divided by $mb$, and it is what makes the
heavy-fuselage limit finite. Written in the wing's normalisation the fuselage inertia is
$1/\MR$, which is infinite in the limit the whole of this paper rests on. Written in its own
it is $\MR + 1$, and $\MR = 0$ gives $y'' = 0$ exactly.

\textbf{The stiffness is padded.} The springs and dampers act across the suspension and feel
the relative displacement alone. The potential energy does not contain $y$, so $\vect{K}$,
$\vect{C}$ and the polynomial springs are the companion paper's with a zero row and a zero
column appended, with no map at all. The restoring forces are
\begin{equation}
\vect{f}_{\mathrm{nl}}(\vect{q}) =
\begin{bmatrix}
(\omega/\Us)^{2}\left(\beta_{\xi}\xi^{3}+\beta_{\xi5}\xi^{5}\right) \\[2pt]
(1/\Us)^{2}\left(\beta_{\alpha}\alpha^{3}+\beta_{\alpha5}\alpha^{5}\right) \\[2pt]
(\omega_{\delta}/\Us)^{2}\left(\beta_{\delta}\delta^{3}+\beta_{\delta5}\delta^{5}\right) \\[2pt]
0
\end{bmatrix}.
\label{eq:nonlin}
\end{equation}
The quintic terms are zero throughout and only the pitch spring is nonlinear, with
$\beta_\alpha = 100$ holding the cycle below five degrees of pitch, inside the attached-flow
validity of the aerodynamics.

Everything the support degree of freedom does follows from that asymmetry between the two
facts. Subtracting the plunge equation from the fuselage equation, every aerodynamic term
cancels, the fuselage row of each aerodynamic operator being $\MR$ times the plunge row, and
the inertia rows differ by a single entry. What is left is
\begin{equation}
y'' = \MR\left[(\omega/\Us)^{2}\left(\xi + \beta_\xi \xi^{3} + \beta_{\xi5}\xi^{5}\right)
+ 2\zeta_\xi(\omega/\Us)\,\xi'\right],
\label{eq:fuselage}
\end{equation}
the statement that the fuselage is driven by the suspension's reaction and by nothing else. The
parameters are the companion paper's, $\mu = 100$, $\omega = 0.2$, $\omega_\delta = 3.5$,
$a_h = -0.5$, $x_\alpha = 0.25$, $r_\alpha = 0.5$, $c = 0.5$, $r_\delta = 0.0791$ and
$x_\delta = 0.0125$, with no structural damping. Only $\MR$ is new.

\subsection{Unsteady aerodynamics}
\label{sec:aero}

The aerodynamics are the companion paper's strip theory in indicial form, with Wagner's and
K\"ussner's functions in the usual two-exponential fit,
\begin{equation*}
\phi(\tau) = 1-\psi_1 e^{-\varepsilon_1\tau}-\psi_2 e^{-\varepsilon_2\tau},
\qquad
\psi(\tau) = 1-\psi_3 e^{-\varepsilon_3\tau}-\psi_4 e^{-\varepsilon_4\tau},
\end{equation*}
with $\psi_1=0.165$, $\psi_2=0.335$, $\varepsilon_1=0.0455$, $\varepsilon_2=0.300$ and
$\psi_3=0.5792$, $\psi_4=0.4208$, $\varepsilon_3=0.1393$, $\varepsilon_4=1.802$, following the
pitch-plunge-flap formulation of Lee and co-workers~\citep{lee1999} as implemented
in~\citep{tantaroudas2015phd}. Six lag states carry the motion and two the gust,
\begin{equation}
w_{i}' = (\vect{T}\vect{q})_{k(i)} - \varepsilon_{\sigma(i)}\,w_{i},
\qquad
w_{7}' = -\varepsilon_{3}w_{7} + w_{g},
\qquad
w_{8}' = -\varepsilon_{4}w_{8} + w_{g},
\label{eq:lag}
\end{equation}
with $w_1,w_2$ driven by $\alpha$, $w_3,w_4$ by the \emph{absolute} plunge $\xi + y$ and
$w_5,w_6$ by $\delta$. The gust velocity is $w_g = 0$ throughout and the two K\"ussner states
are inert in every result reported here, kept so that the model is the solver's own layout.

The driving of $w_3$ and $w_4$ by $\xi + y$ rather than by $\xi$ is the whole of the
aerodynamic change, and it has a consequence worth stating. A displacement that leaves
$\xi + y$ unchanged produces no aerodynamic load whatever, so the section is loaded by its
motion through the air and not by its position in the laboratory, as it must be.

\subsection{Piezoelectric transduction}
\label{sec:piezo}

The transducer is the companion paper's, a lumped element across one structural degree of
freedom with linear electromechanical coupling $\theta$ and capacitance $C_p$, shunted by a
resistive load $R$. Scaling its voltage as $v = C_pV/(\theta b)$, the charge balance is a
first-order equation driven by the velocity of the mounted degree of freedom,
\begin{equation}
v' = q_{j}' - \lambda v,
\qquad
\lambda = \frac{1}{\Us r},
\qquad
\chi = \frac{\theta^{2}}{k_{j} C_{p}},
\qquad
r = R\,C_{p}\,\omega_{\alpha},
\qquad
\kappa = \chi\,k_{j},
\label{eq:circuit}
\end{equation}
with $k_j$ the linear stiffness of the mounted degree of freedom, equal to $(\omega/\Us)^2$ for
plunge, $1/\Us{}^2$ for pitch and $(\omega_\delta/\Us)^2$ for flap. The instantaneous power in
the load and its average over a whole number of cycles are
\begin{equation}
\bar{p} = \frac{\chi v^{2}}{r},
\qquad
\langle\bar{p}\rangle = \frac{1}{T}\int_{0}^{T}\frac{\chi v^{2}}{r}\,\mathrm{d}\tau,
\label{eq:power}
\end{equation}
and because Eq.~\eqref{eq:power} is normalised by $k_j$, powers from different mountings are
referred to the common reference $P/(mb^2\omega_\alpha^3)$ by multiplying by $\omega^2$ for
plunge, $r_\alpha^2$ for pitch or $r_\delta^2\omega_\delta^2$ for flap.

There are three mountings and there cannot be a fourth. A transducer harvests across a spring
and the fuselage has none, being free with nothing to react against. The wing-to-fuselage
transducer \emph{is} the plunge mounting, since $\xi$ is already the relative coordinate, which
is why the plunge mounting is the one the support mass acts on most directly. Its reaction
enters the plunge equation alone and the fuselage feels it through that equation exactly as it
feels the plunge spring.

Collecting $\vect{x} = (\vect{q},\vect{q}',w_1,\dots,w_8,v)$ gives the coupled system as a
seventeen-state autonomous problem, sixteen without the transducer,
\begin{equation}
\vect{x}' = \vect{F}(\vect{x};\Us) = \vect{A}(\Us)\vect{x} + \vect{f}(\vect{x};\Us),
\qquad \vect{x}\in\mathbb{R}^{17}.
\label{eq:fom}
\end{equation}

\subsection{What a free support does to the spectrum}
\label{sec:freefree}

One eigenvalue of $\vect{A}(\Us)$ is an exact algebraic zero at every flow speed. The position
of a free-free system is indifferent, and in the indicial aerodynamic model a steady
displacement carries no load at all, the aerodynamic stiffness being balanced by the lag states
it has driven to their steady values. The null vector is therefore known in closed form,
\begin{equation}
\vect{n} = \left(\vect{e}_y,\ \vect{0},\ -\vect{E}^{-1}\vect{S}\vect{e}_y,\ 0\right),
\label{eq:null}
\end{equation}
with $\vect{E}$ and $\vect{S}$ the lag-state decay and driving matrices of
Eq.~\eqref{eq:lag}. Its partner, the fuselage's \emph{velocity}, is not neutral. Translating
steadily makes a downwash, the air damps it, and the mode sits a little way into the left half
plane, real when the fuselage is heavy and a complex pair once it is light enough to coalesce
with the plunge mode. At $\MR = 0$ the fuselage coordinate is frozen rather than free and
contributes two zeros, decoupled from everything.

Nothing that measures stability may take $\max \mathrm{Re}\,\lambda$ at face value on this
system, and the consequences of that are not confined to bookkeeping;
Section~\ref{sec:impl} records what it cost.

% ============================================================
\section{Implementation and validation}
\label{sec:impl}

The solver is the companion paper's, extended from three degrees of freedom to four by the two
maps of Eq.~\eqref{eq:Tmap}. The aerodynamic algebra is not retyped, it is mapped, so the
$\MR \to 0$ limit is exact by construction rather than by agreement, and the section-frame
operators are held identically equal to the companion paper's by test.

\subsection{Against the solver this one came from}

The four-degree-of-freedom section with a free support was implemented independently in 2013 as
a two-degree-of-freedom pitch-plunge aerofoil on a free fuselage with a massless prescribed
flap. That program is a different one in every respect that matters. It fuses the structural
and aerodynamic terms into some sixty scalars, it orders its state differently, and it carries
the wing's plunge and the fuselage's translation through \emph{separate} Wagner lag states which
its loads then add together, where the present solver carries the single state driven by
$\xi + y$. Agreement between the two is therefore evidence about the physics and not about a
shared subroutine.

The comparison is made on the system both contain, by deleting the flap's row and column from
the assembled operators before the mass matrix is inverted. Over four hundred reduced time
units from the same initial pitch, linear and with the cubic pitch spring active, the
$(\alpha,\xi,y)$ trajectories agree to $1.0\times10^{-14}$ relative. Every eigenvalue of the
present solver's flap-deleted system matches an eigenvalue of the 2013 one to
$10^{-15}$, at three reduced velocities and three support masses, and the 2013 solver's flutter
boundary on its own deck, $\Us_F = 2.180076$ at frequency $0.4341$, is reproduced to
$10^{-6}$.

\subsection{Five independent checks}

\begin{figure}[H]
\centering
\includegraphics[width=\textwidth]{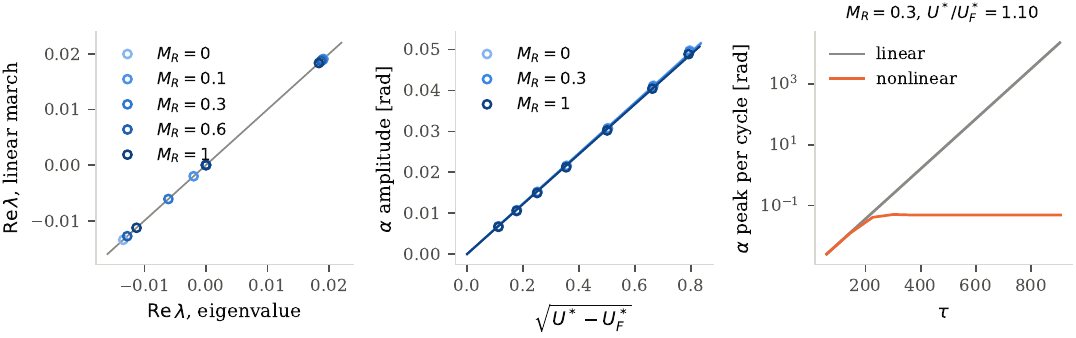}
\caption{Validation. The growth rate the linear march shows against the one the eigenvalue
problem reports (left), the limit-cycle amplitude against the square root of the excess
velocity (centre), and the peak per cycle of the linear and nonlinear marches above the
boundary (right).}
\label{fig:validation}
\end{figure}

Each of the checks in Table~\ref{tab:validation} measures the same system a different way, so
that agreement between them is evidence rather than a repeated calculation. They are plotted
in Figure~\ref{fig:validation}.

\begin{table}[H]
\centering
\caption{The five validation checks and the quantity each one bounds. The worst case over
every support mass and reduced velocity tested is reported.}
\label{tab:validation}
\small
%% generated by scripts/tables.py from data/, do not edit
\begin{tabular}{llr}
\toprule
check & quantity bounded & value \\
\midrule
V1 \enskip growth rate, eigenvalue against linear march & $|\Delta\,\mathrm{Re}\,\lambda|/|\lambda|$ & 4.3e-07 \\
V1 \enskip frequency, eigenvalue against linear march & $|\Delta\,\mathrm{Im}\,\lambda|/|\lambda|$ & 8.4e-06 \\
V2 \enskip heavy-fuselage limit against the companion paper & $|\Delta \Us_F|$ & 5.8e-08 \\
V3 \enskip square-root law above the bifurcation & largest residual & 1.2e-02 \\
V4 \enskip linear against nonlinear below the boundary & largest relative difference & 2.4e-04 \\
V5 \enskip energy balance over five whole cycles & $|W_a - W_e|/W_e$ & 2.9e-03 \\
\bottomrule
\end{tabular}

\end{table}

\textbf{V1, the eigenvalue against the linear march.} The linear response is projected on each
mode's left eigenvector, which annihilates every other mode exactly, rather than having the
envelope of a coordinate fitted, which at a speed where two modes are similarly damped would
fit a beat and not a mode. The recovered growth rates agree with $\mathrm{Re}\,\lambda$ to
$4.3\times10^{-7}$ of $|\lambda|$ and the frequencies with $\mathrm{Im}\,\lambda$ to
$8.4\times10^{-6}$.

\textbf{V2, the heavy-fuselage limit.} At $\MR = 0$ the flutter boundary is $6.3245919$ against
the companion paper's $6.324592$, and its frequency $0.084552$ against that paper's $0.084552$.
The boundary leaves that value smoothly as the support is softened, by $-0.010\%$ at
$\MR = 0.01$ and $-1.168\%$ at $\MR = 1$, with no jump where the fuselage equation changes
character.

\textbf{V3, the supercritical Hopf.} Above the boundary the limit-cycle amplitude is
proportional to the square root of the excess velocity, which is simultaneously a statement
about the nonlinear march and a check that the boundary the eigenvalue problem reports is the
one the march bifurcates at. A straight line through the origin fits the pitch amplitude to
within $1.2\%$ over a ten per cent range above the boundary, at $\MR = 0$, $0.3$ and $1$, and
the march decays to the origin at $0.998\,\UsF$.

\textbf{V4, linear against nonlinear.} The two marches differ in the polynomial springs and in
nothing else, sharing the operator and the step. Below the boundary they agree to
$2.4\times10^{-4}$ while the amplitude is small. Above it the linear march reaches
$6.5\times10^{4}$ rad where the nonlinear one settles at $0.0517$, and the cycle's period is
$75.69$ against the linear $84.51$, the nonlinearity stiffening the pitch spring.

\textbf{V5, the energy balance.} Over five whole cycles the work the air does equals what the
load dissipates to between $7.1\times10^{-4}$ and $2.9\times10^{-3}$ of itself, depending on
the support mass, and the gap shrinks as the cycle is converged further, which is what
identifies it as the cycle not being closed rather than an error in the balance. Stating that balance at all required undoing the equation normalisations, each
equation being divided by its own scale so that a residual dotted with a velocity is not a
power. The weights that undo them are $(1, r_\alpha^2, r_\delta^2, 1/\MR)$, identified by every
conservative operator of the system being symmetric under them, and the fuselage's apparently
singular weight cancels against the $\MR$ in its aerodynamic row to leave lift times
\emph{absolute} plunge rate.

\subsection{What the free-free mode cost}
\label{sec:leading}

Two defects were found in the course of this work, both of them consequences of
Section~\ref{sec:freefree} and both of them silent, and they are recorded because a reader
implementing the same model will meet them.

The first is that a flutter search which bisects on $\max\mathrm{Re}\,\lambda$ does not
terminate meaningfully on a free-free system. That quantity is zero at every flow speed, never
changes sign, and the search converges on the rigid-body mode and reports it, at zero
frequency, as the flutter boundary.

The obvious repair is to exclude eigenvalues within some tolerance of the origin, and it is not
safe. The zero is nearly defective here, its partner being close to it, so a dense solve
returns it at around $10^{-9}$ rather than at machine precision and with either sign. A
transducer on a nearly open circuit has a genuine pole at $-1/(\Us r)$, which at $r = 10^{6}$
is $1.6\times10^{-7}$. No tolerance on the modulus separates $10^{-9}$ of numerical noise from
$1.6\times10^{-7}$ of physics by a safe margin, and one that was tried put the flap mounting's
open-circuit boundary at $\Us = 3.66$ instead of $6.24$. The \emph{eigenvector} separates them
completely, the rigid-body mode being the pure translation of Eq.~\eqref{eq:null} and the
electrical mode being almost entirely voltage, and that is the test the solver now applies.
Equation~\eqref{eq:null} satisfies $\vect{A}\vect{n} = \vect{0}$ to $1.7\times10^{-18}$.

The second is that the limit cycle of a free-free system does not start with its centre of mass
at rest. The transient that grows the cycle carries a net impulse which the neutral mode
integrates, so the cycle sits on an offset and drifts until the damped velocity mode has
decayed. The fuselage amplitude is therefore measured about the mean of its own record rather
than about the origin, and the centre of mass is checked cycle by cycle rather than over a
window of fixed length, a window that slices cycles reporting their own oscillation as drift.

% ============================================================
\section{The flutter boundary}
\label{sec:flutter}

\subsection{The support mass on its own}

\begin{figure}[H]
\centering
\includegraphics[width=\textwidth]{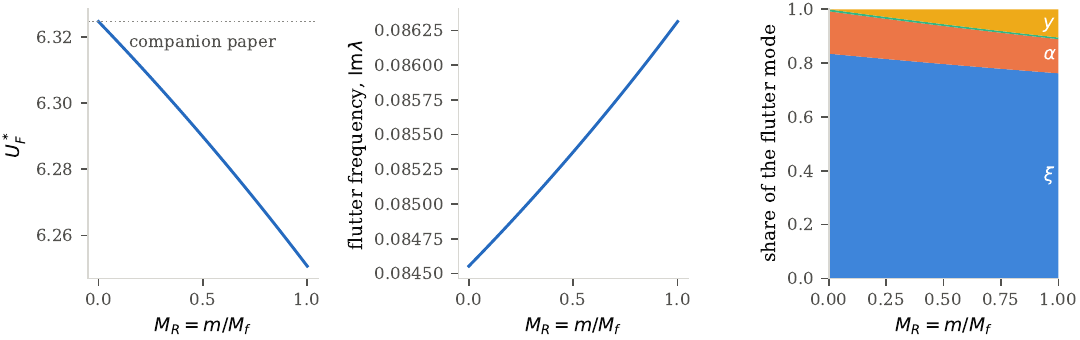}
\caption{The flutter boundary (left), its frequency (centre) and the share of the flutter mode
carried by the plunge, the pitch and the support (right), against the support mass.}
\label{fig:boundary}
\end{figure}

A free support is destabilising, monotonically and mildly (Table~\ref{tab:boundary},
Figure~\ref{fig:boundary}). The boundary falls from $6.3246$ at $\MR = 0$ to $6.2507$ at
$\MR = 1$, a shift of $-1.168\%$, and the flutter frequency rises from $0.084552$ to
$0.086315$.

The mechanism is visible in the mode. The fuselage takes a growing share of the flutter
eigenvector, $10.5\%$ at $\MR = 1$ on the common mechanical reference, and it moves in
antiphase with the wing's plunge. The wing therefore reaches a given absolute plunge with less
compression of its suspension than it would on a rigid mounting, the restoring force at a given
aerodynamic amplitude is smaller, and the mode is softer. The same softening shows in the limit
cycle, whose plunge-to-pitch ratio grows from $2.66$ to $2.97$ over the same range.

\begin{table}[H]
\centering
\caption{The flutter boundary, its frequency and the share of the flutter mode carried by the
plunge, the pitch and the support, against the support mass. The flap carries the small
remainder, $0.0074$ at $\MR = 0$ falling to $0.0062$ at $\MR = 1$. The $\MR = 0$ row is the
companion paper's.}
\label{tab:boundary}
\small
%% generated by scripts/tables.py from data/, do not edit
\begin{tabular}{rrrrrrrr}
\toprule
$M_R$ & $\omega_y/\omega_\alpha$ & $\Us_F$ & shift [\%] & $\mathrm{Im}\,\lambda$ & \multicolumn{3}{c}{share of the flutter mode} \\
\cmidrule(lr){6-8}
 & & & & & $\xi$ & $\alpha$ & $y$ \\
\midrule
0.00 & 0.0000 & 6.3246 & +0.000 & 0.084552 & 0.835 & 0.157 & 0.000 \\
0.10 & 0.0632 & 6.3179 & -0.105 & 0.084708 & 0.827 & 0.154 & 0.012 \\
0.20 & 0.0894 & 6.3111 & -0.213 & 0.084868 & 0.819 & 0.151 & 0.023 \\
0.30 & 0.1095 & 6.3042 & -0.323 & 0.085033 & 0.812 & 0.148 & 0.034 \\
0.50 & 0.1414 & 6.2898 & -0.550 & 0.085375 & 0.797 & 0.141 & 0.055 \\
0.70 & 0.1673 & 6.2747 & -0.789 & 0.085736 & 0.782 & 0.135 & 0.076 \\
1.00 & 0.2000 & 6.2507 & -1.168 & 0.086315 & 0.762 & 0.127 & 0.105 \\
\bottomrule
\end{tabular}

\end{table}

The shift is small in absolute terms, and it is worth putting in the units a harvester designer
already has. A plunge-mounted transducer at $r = 10$ moves the boundary by the same amount as
$\MR = 1$ does when its coupling is $\chi = 0.17$, which is a third of the strongest coupling
the companion paper considered. A support mass equal to the wing section is therefore a
first-order effect on the same scale as the transducer itself, not a correction to it.

\subsection{The transducer on a free support}

\begin{figure}[H]
\centering
\includegraphics[width=\textwidth]{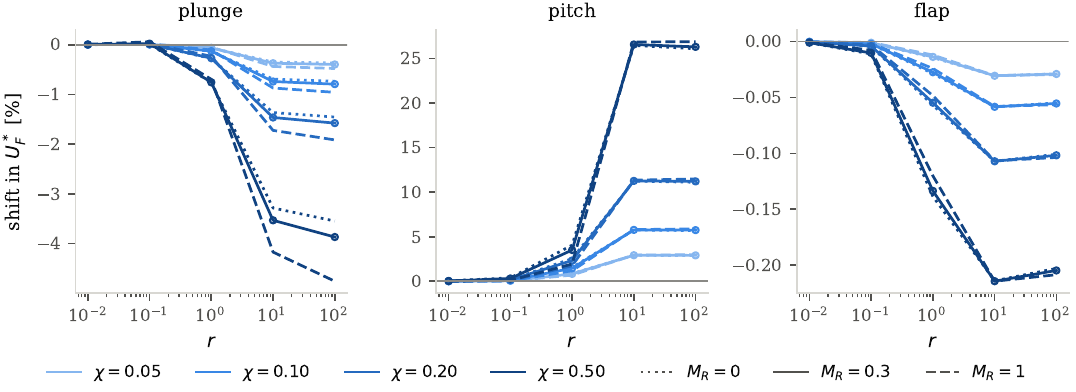}
\caption{Shift in the flutter boundary against load resistance, measured from each support
mass's own baseline, one panel per mounting.}
\label{fig:flutter}
\end{figure}

Table~\ref{tab:flutter_shift} is the companion paper's Table~2 extended by the support mass,
and Figure~\ref{fig:flutter} plots it. The $\MR = 0$ block reproduces that paper's published
table to $4.8\times10^{-5}$, which is its own rounding, over all forty entries, and the
$\MR > 0$ blocks are the new result. Every shift is measured from the boundary of the \emph{same}
support mass, so that the transducer's effect and the support's are separated rather than
added.

\begin{table}[H]
\centering
\caption{Linear flutter boundary against load resistance for the three mountings at four
couplings, at three support masses. The $\MR = 0$ block reproduces the companion paper.}
\label{tab:flutter_shift}
\small
%% generated by scripts/tables.py from data/, do not edit
\begin{tabular}{llrrrrr}
\toprule
mounting & $\chi$ & \multicolumn{5}{c}{$\Us_F$ at load resistance $r$} \\
\cmidrule(lr){3-7}
 & & $r = 0.01$ & $r = 0.1$ & $r = 1$ & $r = 10$ & $r = 100$ \\
\midrule
\multicolumn{7}{l}{\itshape $M_R = 0$, baseline $\Us_F = 6.3246$} \\
plunge & 0.05 & 6.3246 & 6.3247 & 6.3204 & 6.3026 & 6.3013 \\
plunge & 0.10 & 6.3246 & 6.3247 & 6.3161 & 6.2808 & 6.2781 \\
plunge & 0.20 & 6.3246 & 6.3249 & 6.3069 & 6.2382 & 6.2325 \\
plunge & 0.50 & 6.3247 & 6.3252 & 6.2763 & 6.1168 & 6.1002 \\
\addlinespace
pitch & 0.05 & 6.3249 & 6.3279 & 6.3801 & 6.5090 & 6.5067 \\
pitch & 0.10 & 6.3252 & 6.3310 & 6.4232 & 6.6886 & 6.6843 \\
pitch & 0.20 & 6.3258 & 6.3365 & 6.4850 & 7.0352 & 7.0274 \\
pitch & 0.50 & 6.3274 & 6.3489 & 6.5812 & 7.9987 & 7.9753 \\
\addlinespace
flap & 0.05 & 6.3246 & 6.3245 & 6.3237 & 6.3227 & 6.3228 \\
flap & 0.10 & 6.3246 & 6.3245 & 6.3228 & 6.3209 & 6.3211 \\
flap & 0.20 & 6.3246 & 6.3243 & 6.3210 & 6.3178 & 6.3182 \\
flap & 0.50 & 6.3245 & 6.3239 & 6.3158 & 6.3111 & 6.3118 \\
\midrule
\multicolumn{7}{l}{\itshape $M_R = 0.3$, baseline $\Us_F = 6.3042$} \\
plunge & 0.05 & 6.3042 & 6.3043 & 6.3002 & 6.2807 & 6.2791 \\
plunge & 0.10 & 6.3042 & 6.3045 & 6.2960 & 6.2574 & 6.2541 \\
plunge & 0.20 & 6.3043 & 6.3048 & 6.2870 & 6.2118 & 6.2047 \\
plunge & 0.50 & 6.3044 & 6.3055 & 6.2565 & 6.0814 & 6.0602 \\
\addlinespace
pitch & 0.05 & 6.3044 & 6.3070 & 6.3565 & 6.4883 & 6.4874 \\
pitch & 0.10 & 6.3047 & 6.3097 & 6.3962 & 6.6677 & 6.6660 \\
pitch & 0.20 & 6.3052 & 6.3143 & 6.4509 & 7.0144 & 7.0109 \\
pitch & 0.50 & 6.3066 & 6.3238 & 6.5253 & 7.9795 & 7.9629 \\
\addlinespace
flap & 0.05 & 6.3042 & 6.3041 & 6.3033 & 6.3023 & 6.3024 \\
flap & 0.10 & 6.3042 & 6.3041 & 6.3025 & 6.3005 & 6.3007 \\
flap & 0.20 & 6.3042 & 6.3039 & 6.3008 & 6.2974 & 6.2978 \\
flap & 0.50 & 6.3041 & 6.3036 & 6.2958 & 6.2907 & 6.2913 \\
\midrule
\multicolumn{7}{l}{\itshape $M_R = 1$, baseline $\Us_F = 6.2507$} \\
plunge & 0.05 & 6.2507 & 6.2511 & 6.2475 & 6.2233 & 6.2207 \\
plunge & 0.10 & 6.2508 & 6.2515 & 6.2440 & 6.1962 & 6.1907 \\
plunge & 0.20 & 6.2509 & 6.2522 & 6.2361 & 6.1429 & 6.1309 \\
plunge & 0.50 & 6.2512 & 6.2544 & 6.2071 & 5.9899 & 5.9532 \\
\addlinespace
pitch & 0.05 & 6.2508 & 6.2523 & 6.2939 & 6.4342 & 6.4369 \\
pitch & 0.10 & 6.2509 & 6.2536 & 6.3242 & 6.6133 & 6.6183 \\
pitch & 0.20 & 6.2512 & 6.2556 & 6.3594 & 6.9601 & 6.9682 \\
pitch & 0.50 & 6.2518 & 6.2572 & 6.3717 & 7.9296 & 7.9313 \\
\addlinespace
flap & 0.05 & 6.2507 & 6.2506 & 6.2499 & 6.2488 & 6.2488 \\
flap & 0.10 & 6.2507 & 6.2506 & 6.2492 & 6.2471 & 6.2472 \\
flap & 0.20 & 6.2507 & 6.2505 & 6.2477 & 6.2440 & 6.2442 \\
flap & 0.50 & 6.2506 & 6.2502 & 6.2432 & 6.2373 & 6.2377 \\
\bottomrule
\end{tabular}

\end{table}

The three mountings respond to the support differently, and the difference is informative.

\textbf{Plunge, amplified.} The plunge mounting is destabilising on a rigid support and more so
on a free one. At $\chi = 0.50$ in the open-circuit limit the shift grows from $-3.57\%$ to
$-3.90\%$ at $\MR = 0.3$ and $-4.83\%$ at $\MR = 1$, a third larger. That is the mounting whose
transducer acts across the very spring the fuselage hangs from, so the two softenings compound.

\textbf{Pitch, unchanged where it is largest and halved where it is not.} The pitch mounting's
largest stabilising shift barely moves, $+25.94\%$ at $\MR = 0$ against $+26.76\%$ at
$\MR = 1$. At $r = 1$, however, where the electrical time constant $1/(\Us r)$ is near the
flutter frequency, the shift falls from $+4.06\%$ to $+1.94\%$, less than half. The optimum
load resistance moves outward with the support, from $r = 14.1$ to $r = 17.8$ at $\chi = 0.50$.
The transducer's effect on stability in the open-circuit limit is an added stiffness, which the
support mass does not touch. Its effect at finite $r$ is a phase-dependent damping, which is
where a changed mode shape and a changed flutter frequency are felt.

\textbf{Flap, untouched.} The flap mounting shifts the boundary by $-0.20\%$ at $\chi = 0.50$
open circuit at every support mass tested. The flap is an order of magnitude weaker than the
other two mountings on a rigid support, and the support mass does not change that.

% ============================================================
\section{The limit cycle and the harvested power}
\label{sec:power}

\subsection{The cycle}

Above the boundary the response settles onto a limit cycle. At the companion paper's anchor
velocity $\Us = 7.0$ every amplitude grows with the support mass and the period shortens
(Table~\ref{tab:lco}). Without a transducer the pitch amplitude grows from $0.0516$ to
$0.0539$~rad and the plunge from $0.1372$ to $0.1601$ over $\MR = 0$ to $1$, while the period
falls from $76.15$ to $75.23$. The fuselage carries $9.0\%$ of the orbit at $\MR = 1$.

\begin{table}[H]
\centering
\caption{The limit cycle at the anchor velocity against the support mass, without a transducer
and with one on the plunge degree of freedom.}
\label{tab:lco}
\small
%% generated by scripts/tables.py from data/, do not edit
\begin{tabular}{rrrrrrrr}
\toprule
$M_R$ & $\Us_F$ & $\Us/\Us_F$ & $\alpha$ [rad] & $\xi$ & $\delta$ [rad] & $y$ & period \\
\midrule
\multicolumn{8}{l}{\itshape no transducer} \\
0.00 & 6.3246 & 1.1068 & 0.0516 & 0.1372 & 0.00636 & 0.0000 & 76.15 \\
0.10 & 6.3179 & 1.1080 & 0.0518 & 0.1393 & 0.00642 & 0.0017 & 76.07 \\
0.20 & 6.3111 & 1.1091 & 0.0520 & 0.1414 & 0.00647 & 0.0034 & 75.98 \\
0.30 & 6.3042 & 1.1104 & 0.0522 & 0.1436 & 0.00653 & 0.0051 & 75.89 \\
0.50 & 6.2898 & 1.1129 & 0.0527 & 0.1480 & 0.00665 & 0.0088 & 75.71 \\
0.70 & 6.2747 & 1.1156 & 0.0532 & 0.1527 & 0.00676 & 0.0126 & 75.52 \\
1.00 & 6.2507 & 1.1199 & 0.0539 & 0.1601 & 0.00694 & 0.0187 & 75.23 \\
\addlinespace
\multicolumn{8}{l}{\itshape $\chi = 0.20$, $r = 1$, plunge} \\
0.00 & 6.3069 & 1.1099 & 0.0524 & 0.1318 & 0.00652 & 0.0000 & 74.32 \\
0.10 & 6.3004 & 1.1110 & 0.0526 & 0.1336 & 0.00658 & 0.0016 & 74.20 \\
0.20 & 6.2938 & 1.1122 & 0.0528 & 0.1354 & 0.00663 & 0.0032 & 74.08 \\
0.30 & 6.2870 & 1.1134 & 0.0530 & 0.1373 & 0.00669 & 0.0049 & 73.95 \\
0.50 & 6.2731 & 1.1159 & 0.0535 & 0.1412 & 0.00682 & 0.0084 & 73.70 \\
0.70 & 6.2587 & 1.1184 & 0.0540 & 0.1451 & 0.00695 & 0.0120 & 73.45 \\
1.00 & 6.2361 & 1.1225 & 0.0547 & 0.1513 & 0.00716 & 0.0176 & 73.05 \\
\bottomrule
\end{tabular}

\end{table}

\subsection{The coupling and load-resistance plane}

\begin{figure}[H]
\centering
\includegraphics[width=0.96\textwidth]{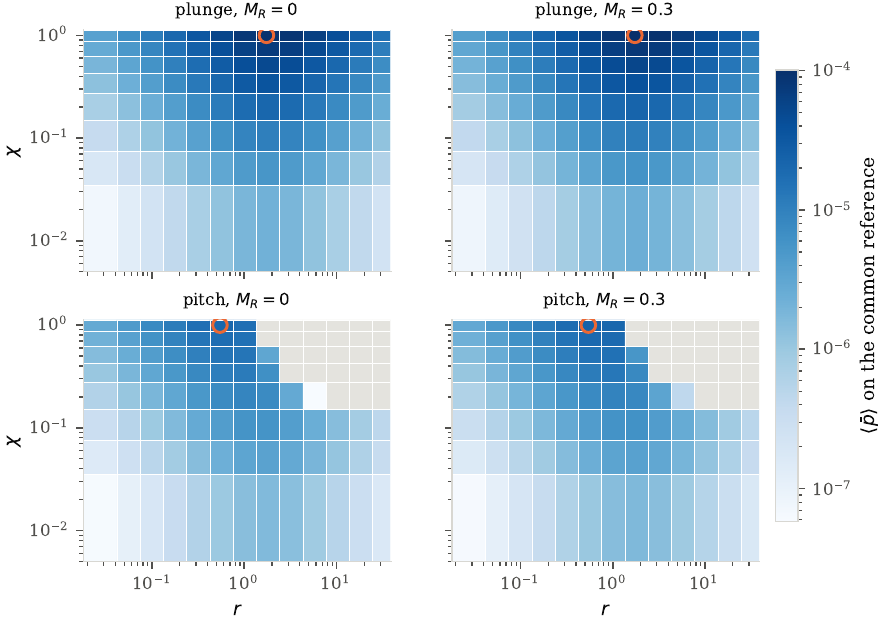}
\caption{Mean harvested power over the coupling and load-resistance plane at the anchor
velocity, on the common mechanical reference and on one logarithmic scale for all four panels.
Grey cells have no converged limit cycle and the circle marks each optimum.}
\label{fig:powermap}
\end{figure}

The mean power over the $(\chi,r)$ plane at the anchor velocity is shown in
Figure~\ref{fig:powermap} and its optima in Table~\ref{tab:power}. The $\MR = 0$ maps reproduce
the companion paper. The plunge optimum is $\chi = 1.00$, $r = 1.74$ at
$9.335\times10^{-5}$ against that paper's $9.35\times10^{-5}$ at $r = 1.78$, the pitch optimum
$\chi = 1.00$, $r = 0.55$ at $2.080\times10^{-5}$ against its $2.09\times10^{-5}$ at
$r = 0.56$, and $79$ of the $104$ pitch-mounted cells reach a converged cycle against its $78$,
the remainder being linearly stable and supporting none.

\begin{table}[H]
\centering
\caption{The optimum of each power map, against the companion paper's.}
\label{tab:power}
\small
%% generated by scripts/tables.py from data/, do not edit
\begin{tabular}{llrrrr}
\toprule
source & mounting & $\chi$ & $r$ & $\langle\bar p\rangle$ & cells with a cycle \\
\midrule
companion paper & plunge & 1.00 & 1.78 & 9.35e-05 & --- \\
companion paper & pitch & 1.00 & 0.56 & 2.09e-05 & --- \\
\addlinespace
$M_R = 0$ & plunge & 1.00 & 1.74 & 9.335e-05 & 104 of 104 \\
$M_R = 0$ & pitch & 1.00 & 0.55 & 2.080e-05 & 79 of 104 \\
\addlinespace
$M_R = 0.3$ & plunge & 1.00 & 1.74 & 1.012e-04 & 104 of 104 \\
$M_R = 0.3$ & pitch & 1.00 & 0.55 & 2.360e-05 & 79 of 104 \\
\bottomrule
\end{tabular}

\end{table}

The optimum does not move. At $\MR = 0.3$ it is at the same $(\chi, r)$ cell for both
mountings, and the power at it rises to $1.012\times10^{-4}$ and $2.360\times10^{-5}$. The
matching rules of the rigidly supported section therefore carry over to a free one unchanged,
which is the most directly useful finding here. The plunge mounting's advantage over the pitch
mounting narrows slightly, from a factor $4.49$ to $4.29$.

\subsection{Power against the support mass, measured two ways}
\label{sec:twoways}

\begin{figure}[H]
\centering
\includegraphics[width=0.94\textwidth]{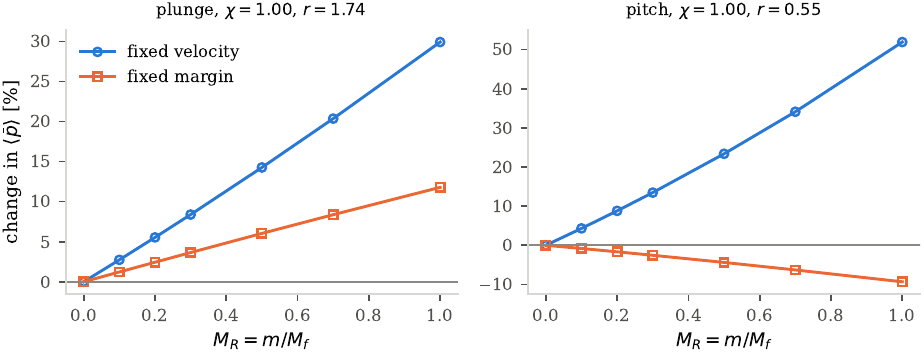}
\caption{Change in the mean harvested power with the support mass at each mounting's optimum,
at a fixed reduced velocity and at a fixed margin above each system's own flutter boundary.}
\label{fig:powervs}
\end{figure}

The headline result had to be measured twice, and the difference between the two measurements
is the result (Figure~\ref{fig:powervs}).

At a \emph{fixed reduced velocity}, which is the convention of the companion paper and of the
maps above, the harvested power rises with the support mass for both mountings, by $+29.9\%$
for the plunge mounting and $+51.9\%$ for the pitch mounting at $\MR = 1$. Read naively that
says a soft support is good for a harvester and better for a pitch-mounted one.

At a \emph{fixed margin} above each system's own flutter boundary, which divides out the fact
that the support mass moves the boundary and asks what it does to the cycle itself, the plunge
mounting keeps $+11.8\%$ and the pitch mounting \emph{loses} $9.3\%$.

The pitch mounting's apparent gain is therefore entirely the boundary coming down to meet the
flow speed, and none of it is a better cycle. The plunge mounting's gain is about two fifths
real and three fifths the same effect. The distinction is not academic. A harvester operates at
a flow speed it is given, so the fixed-velocity number is the one that predicts its output at
that speed. A claim that a softer support produces a better limit cycle, or a comparison
between two harvesters whose boundaries differ, needs the fixed-margin number instead, and the
two differ by the whole of the gain for the pitch mounting.

\begin{figure}[H]
\centering
\includegraphics[width=\textwidth]{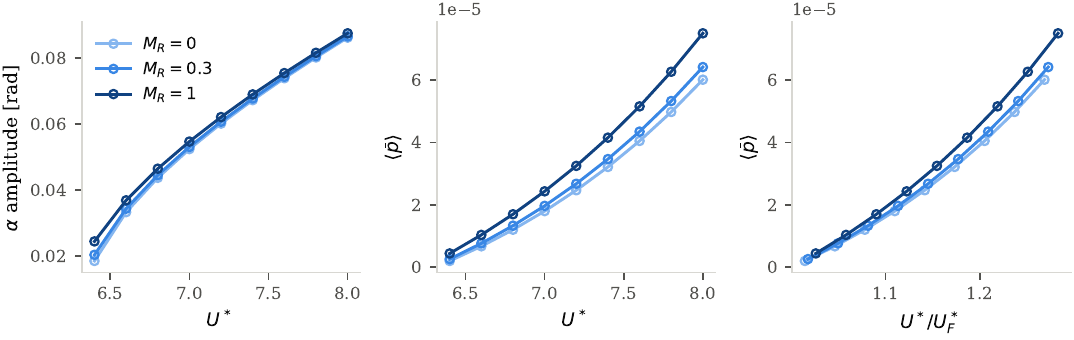}
\caption{Pitch amplitude and mean power against reduced velocity, and the same power against
the margin above each system's own boundary, at a fixed transducer.}
\label{fig:velocity}
\end{figure}

The velocity sweep in Figure~\ref{fig:velocity} shows the same thing directly. Plotted against
$\Us$ the three support masses separate, and plotted against $\Us/\UsF$ they nearly collapse.

% ============================================================
\section{The reduced model}
\label{sec:rom}

The companion paper reduced its fifteen-state harvester to four states by projecting the
residual onto two eigenvector pairs of the coupled Jacobian, and concluded that such a model
must be judged on the harvested quantity rather than on a structural amplitude, its pitch
being accurate while its voltage was not. A free support raises a question that a grounded one
cannot. The system now has a rigid-body mode at the origin and a slow mode beside it, neither
of them oscillatory at the flutter frequency and neither contained in any basis of
flutter-band pairs, so the reduction may be missing a whole part of the motion.

The reduction here is the companion paper's with its operator re-projected rather than
expanded, which is the variant that paper found correct. With $\vect{\Phi}$ the retained right
eigenvectors of $\vect{A}(\Us_{\mathrm{ref}})$ and $\vect{\Psi}$ the left ones, the real basis
is $\vect{V} = [\,2\,\mathrm{Re}\,\vect{\Phi},\ -2\,\mathrm{Im}\,\vect{\Phi}\,]$ and
$\vect{W} = [\,\mathrm{Re}\,\vect{\Psi},\ -\mathrm{Im}\,\vect{\Psi}\,]$, biorthonormalised so
that $\vect{W}^{\mathsf T}\vect{V} = \vect{I}$, a real mode contributing one column rather than
two. The reduced residual is $\vect{W}^{\mathsf T}\vect{F}(\vect{V}\vect{z})$, whose linear
part $\vect{W}^{\mathsf T}\vect{A}(\Us)\vect{V}$ is evaluated at the running velocity on the
frozen basis, and whose cubic spring is projected directly. Three bases are compared, the
companion paper's two structural pairs, the flutter pair with the two rigid-body modes in
place of the second structural pair, and all of them together.

\begin{table}[H]
\centering
\caption{Relative error of the reduced model against the full one at the anchor velocity, per
cent, for three bases and a ladder of support masses.}
\label{tab:rom}
\small
%% generated by scripts/tables.py from data/, do not edit
\begin{tabular}{rrrrrrrr}
\toprule
$M_R$ & states & $\alpha$ & $\xi$ & $y$ & $v$ & $\langle\bar p\rangle$ & period \\
\midrule
\multicolumn{8}{l}{\itshape two structural pairs, the companion paper's rule} \\
0.00 & 4 & +0.1 & +0.8 & --- & -21.9 & -27.9 & -2.0 \\
0.10 & 4 & +0.2 & -0.1 & +28.2 & -23.1 & -29.8 & -2.2 \\
0.20 & 4 & +0.3 & -1.1 & +27.4 & -24.3 & -31.8 & -2.4 \\
0.30 & 4 & +0.3 & -2.1 & +26.6 & -25.6 & -33.8 & -2.6 \\
0.50 & 4 & +0.5 & -4.1 & +25.0 & -28.2 & -37.9 & -3.1 \\
0.70 & 4 & +0.6 & -6.3 & +23.4 & -30.9 & -42.0 & -3.7 \\
1.00 & 4 & +0.9 & -9.8 & +21.4 & -35.3 & -48.2 & -4.5 \\
\addlinespace
\multicolumn{8}{l}{\itshape flutter pair and the rigid-body modes} \\
0.10 & 4 & +11.8 & +78.6 & +710.8 & +26.7 & +65.7 & +1.0 \\
0.20 & 4 & +12.1 & +89.5 & +1048.1 & +25.0 & +61.4 & +0.7 \\
0.30 & 4 & +12.7 & +121.5 & +2272.2 & +17.8 & +42.5 & -0.2 \\
0.50 & 5 & +12.5 & +63.3 & +115.3 & +56.6 & +161.4 & +2.5 \\
0.70 & 5 & +13.0 & +66.4 & +119.3 & +60.1 & +173.6 & +2.5 \\
1.00 & 5 & +13.8 & +71.6 & +126.0 & +65.8 & +194.5 & +2.5 \\
\addlinespace
\multicolumn{8}{l}{\itshape both structural pairs and the rigid-body modes} \\
0.10 & 6 & +0.2 & +3.5 & +398.8 & -23.4 & -30.2 & -2.3 \\
0.20 & 6 & +0.4 & +8.9 & +623.6 & -25.7 & -33.7 & -2.7 \\
0.30 & 6 & +1.2 & +27.2 & +1437.0 & -32.2 & -42.2 & -3.9 \\
0.50 & 7 & -0.4 & -4.3 & -4.9 & -8.2 & -8.4 & -0.4 \\
0.70 & 7 & -0.5 & -4.4 & -5.0 & -8.3 & -8.8 & -0.4 \\
1.00 & 7 & -0.5 & -4.5 & -5.1 & -8.5 & -9.5 & -0.4 \\
\bottomrule
\end{tabular}

\end{table}

\begin{figure}[H]
\centering
\includegraphics[width=\textwidth]{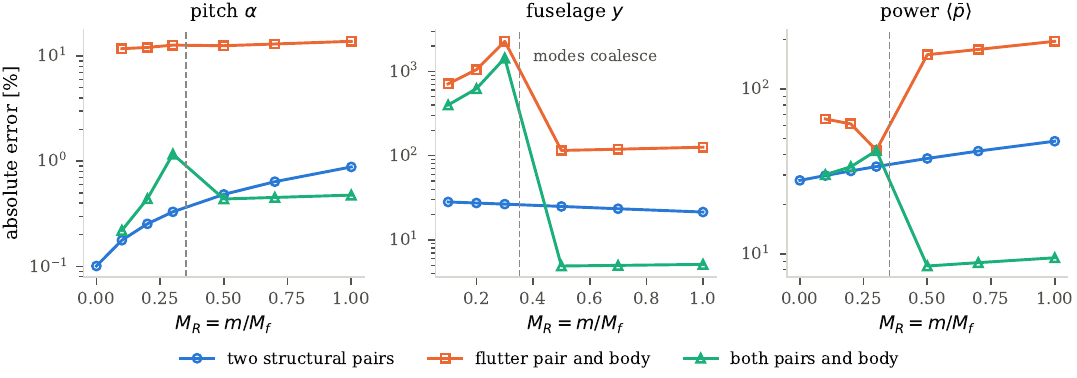}
\caption{Absolute error of the reduced model in the pitch, the fuselage motion and the
harvested power, against the support mass. The dashed line is where the two slow rigid-body
modes coalesce.}
\label{fig:rom}
\end{figure}

\textbf{The companion paper's conclusion is strengthened.} On the two-pair basis the pitch is
reproduced to better than one per cent at every support mass, and the harvested power is not.
The power error grows from $-27.9\%$ on a rigid support to $-48.2\%$ at $\MR = 1$ and the
voltage error from $-21.9\%$ to $-35.3\%$, while the pitch error over the same range goes from
$0.1\%$ to $0.9\%$. A reduced model of this kind on a free support can therefore be wrong about
the harvested power by nearly a factor of two while appearing, on the quantity most naturally
plotted, to be accurate to within one per cent.

\textbf{The fuselage is the part the flutter-band basis misses.} The two-pair basis
misstates the fuselage amplitude by about a quarter at every support mass, $+28.2\%$ at
$\MR = 0.1$ falling slowly to $+21.4\%$ at $\MR = 1$, an error an order of magnitude larger
than the pitch's and one that does not improve as the support is softened and the fuselage
becomes more energetic.

\textbf{Whether retaining the rigid-body modes helps depends on a modal coalescence.} Adding
them is not a refinement of the basis, it is a change of subspace, and the result divides
sharply at $\MR = 0.3519$. Below that value the two slow modes are separate and real, one
carrying the fuselage and one the wing's slow plunge relaxation, and a basis that retains the
neutral mode and one of them leaves the other out. The fuselage amplitude is then worse than
with no rigid-body modes at all, by an order of magnitude, and the power is no better. Above
it the two have merged into a single damped pair which shares the motion between the wing and
the fuselage equally, retaining it captures the whole slow subspace, and every error falls at
once. At $\MR = 1$ the six-state model is within $0.5\%$ on the pitch, $5.1\%$ on the
fuselage, $8.5\%$ on the voltage, $9.5\%$ on the power and $0.4\%$ on the period, against the
four-state model's $48.2\%$ on the power.

\textbf{The second structural pair is not optional.} Spending the same four states on the
flutter pair and the rigid-body modes instead is worse on every quantity at every support
mass, the pitch by more than ten per cent and the power by between forty and two hundred. The
companion paper's basis rule is the right starting point, and the rigid-body modes are an
addition to it and not a substitute.

At $\MR = 0$ the rigid-body modes cannot be retained at all. The support coordinate is frozen
and decoupled, its two eigenvalues are a repeated zero whose eigenspace the coupled system
never enters, and the biorthonormalisation is singular. The rigid-body modes exist as modes
only when the support is free, which is the sense in which this question has no counterpart in
the companion paper.

% ============================================================
\section{Discussion}
\label{sec:discussion}

Table~\ref{tab:comparison} puts the companion paper's headline quantities beside this paper's
at the anchor support mass.

\begin{table}[H]
\centering
\caption{The companion paper's rigidly supported section against the same section on a free
fuselage, at the anchor support mass.}
\label{tab:comparison}
\footnotesize
%% generated by scripts/tables.py from data/, do not edit
\begin{tabular}{lrr}
\toprule
quantity & companion paper, & this work, \\
 & rigid support, $M_R = 0$ & free support, $M_R = 0.3$ \\
\midrule
baseline flutter boundary $\Us_F$ & 6.3246 & 6.3042 \\
plunge mounting, shift at $\chi = 0.50$, open circuit [\%] & -3.57 & -3.90 \\
pitch mounting, shift at $\chi = 0.50$, open circuit [\%] & +25.94 & +26.16 \\
pitch mounting, shift at $\chi = 0.50$, $r = 1$ [\%] & +4.06 & +3.51 \\
flap mounting, shift at $\chi = 0.50$, open circuit [\%] & -0.20 & -0.20 \\
best plunge-mounted $\langle\bar p\rangle$ & 9.335e-05 & 1.012e-04 \\
its $(\chi, r)$ & (1.00, 1.74) & (1.00, 1.74) \\
best pitch-mounted $\langle\bar p\rangle$ & 2.080e-05 & 2.360e-05 \\
plunge over pitch at the optimum & 4.49 & 4.29 \\
\bottomrule
\end{tabular}

\end{table}

\textbf{What survives.} The ranking of the mountings survives unchanged. The plunge mounting
harvests about four and a half times what the pitch mounting does on the common reference, and
the flap mounting is an order of magnitude below both in its effect on stability. The sign of
each mounting's shift survives, plunge destabilising and pitch stabilising. The position of the
optimum in the coupling and load-resistance plane survives, so a harvester matched on a
wind-tunnel rig is matched on a flying one. The character of the bifurcation survives, the
cycle remaining a supercritical Hopf with a square-root amplitude law at every support mass
tested.

\textbf{What does not.} The magnitude of the plunge mounting's destabilisation grows by a third
at $\MR = 1$, which matters because it is a design margin. The pitch mounting's stabilisation
at intermediate load resistance is more than halved, and its optimum resistance moves outward
by a quarter, so a pitch-mounted transducer tuned for stabilisation on a rigid rig is mistuned
on a free support. And the harvested power's dependence on the support is a statement about the
convention as much as about the physics, which is the one finding here that has no counterpart
in the companion paper because a rigidly supported section has no such parameter to vary.

\textbf{What the support mass is worth.} The whole effect is small in absolute terms. A
fuselage as heavy as the wing section moves the flutter boundary by just over one per cent,
which is about what a plunge-mounted transducer with $\chi = 0.17$ does. For a real aircraft,
where the fuselage is many times the mass of a wing section and $\MR$ is correspondingly small,
the effect on the boundary is a fraction of a per cent and the design conclusions of the
rigidly supported section apply as they stand. The regime where the support matters is the one
where $\MR$ approaches unity, which is a small unmanned vehicle, a gust generator or a
harvester built as a free body rather than as an instrumented wing.

\textbf{Limitations.} The aerodynamics are incompressible strip theory, and the companion CFD
study~\citep{tantaroudas2026cfd} found that strip theory under-predicts the harvested power of
the compressible section and overstates the transducer's effect on stability where the
electrical time constant meets the flutter frequency. That last regime is exactly the one where
the support mass was found here to matter most, and the two effects have the \emph{same} sign. At
$\chi = 0.50$ and $r = 1$ the pitch mounting's stabilisation is $+4.06\%$ with strip theory on
a rigid support, $+1.95\%$ with the Euler aerodynamics on a rigid support, and $+1.94\%$ with
strip theory on a free support of $\MR = 1$. Each halves it on its own. Whether the two
compound to a quarter, overlap, or do something else again is not something either study can
answer alone, and the present work does not answer it. It is, however, cheaper to answer than it looks. A flow
solver coupled to this section needs no four-degree-of-freedom mesh motion, because the wall
displacement conjugate to the fuselage translation is a rigid vertical translation and so is
the one conjugate to the plunge. The two are identical, the flow solver still sees three
coordinates, and the same two maps of Eq.~\eqref{eq:Tmap} carry it to four. A companion study
takes that route. The fuselage here is a point mass in plunge alone, with no pitch freedom and
no second wing, so the model is a free-free section and not a free-free aircraft. The flap's hinge carries no gust term, which is immaterial here
because no gust is applied.

% ============================================================
\section{Conclusions}
\label{sec:conclusions}

\begin{enumerate}[leftmargin=*,itemsep=2pt]
\item A pitch-plunge-flap section sprung from a free fuselage is assembled from the rigidly
supported one by two maps, the air acting on the absolute plunge and the springs on the
relative one. The heavy-fuselage limit is then exact rather than approached, and the
implementation reproduces both the companion paper's published flutter table and its power map
to their own rounding.

\item A free support is destabilising and mildly so, by softening the flutter mode. The
fuselage takes a share of it in antiphase with the wing's plunge, so the suspension is less
compressed at a given aerodynamic amplitude. A fuselage as heavy as the wing section is worth
about the same shift as a plunge-mounted transducer of moderate coupling.

\item The support acts on the three mountings unequally. It amplifies the plunge mounting's
destabilisation by a third, leaves the pitch mounting's largest stabilisation alone while more
than halving it where the electrical time constant meets the flutter frequency, and leaves the
flap mounting untouched.

\item The optimum coupling and load resistance do not move, so a harvester matched on a rigid
mounting stays matched on a free one, and the plunge mounting keeps its advantage over the
pitch mounting.

\item The harvested power rises with the support mass at a fixed flow speed for both mountings,
and that is largely a statement about the convention. Measured at a fixed margin above each
system's own boundary the plunge mounting keeps a real gain and the pitch mounting loses power,
so the pitch mounting's apparent benefit is entirely the boundary moving. A harvesting result
quoted at a fixed speed and read as a property of the limit cycle is wrong by the whole of that
difference, and papers in this field should say which convention they used.

\item A four-state reduced model of the companion paper's kind reproduces the pitch of the
free-free section to within one per cent while misstating the harvested power by up to a factor
of two, and misstates the fuselage motion by about a quarter throughout. Retaining the
rigid-body modes as well fixes all of it, but only above the support mass at which the two slow
modes coalesce into a damped pair. Below that they are separate real modes, a basis that takes
one of them leaves the other out, and the fuselage motion comes out an order of magnitude worse
than with no rigid-body modes at all. Enlarging such a basis is a change of subspace and not a
refinement, and it can make a reduced model worse.

\item A free-free aeroelastic system cannot have its stability measured by the largest real
part of its spectrum, and excluding the rigid-body mode by a tolerance on the modulus is not
safe either, because an almost open-circuited transducer has a genuine pole of the same size.
The rigid-body eigenvector is available in closed form and is the reliable test.
\end{enumerate}

% ============================================================
\section*{Data availability}

All data generated in this study, the scripts that produce every figure and table, and the
source code of the aeroelastic solver are available from the corresponding author upon
reasonable request.

\section*{Funding and competing interests}

The authors received no financial support for this work and declare no competing interests.

\section*{Declaration of generative AI use}

Claude (Anthropic) was used to assist in drafting and editing the text, refining its language
and proofreading it, and in checking the analysis scripts and the consistency of the text with
the computed data. The authors reviewed all content and take full responsibility for it.

% ============================================================

\end{document}